\documentclass[twocolumn,preprintnumbers,nofootinbib,prd]{revtex4}

\usepackage{graphicx}

\usepackage[hypertex]{hyperref}
\newcommand{\vev}[1]{\langle {#1} \rangle}
\newcommand{\lsim}{\lesssim}

\newcommand{\gsim}{\gtrsim}
\newcommand{\beq}{\begin{equation}}
\newcommand{\eeq}{\end{equation}}
\newcommand{\be}{B_\oplus}
\newcommand{\re}{R_\oplus}
\newcommand{\unit}[1]{\,\mathrm{#1}}

\begin{document}

\pagestyle{plain}

\preprint{MADPH-05-1439}

\title{Detecting solar axions using Earth's magnetic field}

\author{Hooman Davoudiasl
}

\author{Patrick Huber
}

\affiliation{Department of Physics, University of Wisconsin,
Madison, WI 53706, USA}


\begin{abstract}

We show that solar axion conversion to photons in the Earth's
magnetosphere can produce an x-ray flux, with average energy
$\vev{\omega} \simeq 4\unit{keV}$, which is
measurable on the dark side of the
Earth. The smallness of the Earth's magnetic field is compensated
by a large magnetized volume.  For axion masses
$m_a \lsim 10^{-4}\unit{eV}$, a low-Earth-orbit
x-ray detector with an effective
area of $10^4\unit{cm}^2$, pointed at the solar core, can probe the
photon-axion coupling down to $10^{-11}\unit{GeV}^{-1}$, in
one year. Thus, the sensitivity of this new approach will be an
order of magnitude beyond current laboratory limits.

\end{abstract}
\maketitle


The existence of weakly interacting light pseudo-scalars is 
well-motivated in particle physics.  For example,  
experimental evidence requires the size of $CP$ violation in
strong interactions, as parameterized by the angle $\theta$, to be
very tiny; $\theta \lsim 10^{-10}$.  However, there is no symmetry
reason within the Standard Model (SM) for such a small $\theta$
angle; this is the strong $CP$ problem.  An elegant solution to
this puzzle was proposed by Peccei and Quinn \cite{Peccei:1977hh},
where a new $U(1)$ symmetry, anomalous under strong interactions,
was proposed.  This $U(1)$ symmetry is assumed to be spontaneously
broken at a scale $f_a$, resulting in a pseudo-scalar Goldstone
boson $a$ \cite{Weinberg:1977ma}, the axion.  Non-perturbative QCD
interactions at the scale $\Lambda_{\rm QCD} \sim 100\unit{MeV}$
generate a potential for the axion, endowing it with a mass $m_a
\sim \Lambda_{\rm QCD}^2/f_a$.  Experimental and observational
bounds have pushed $f_a$ to scales of order $10^7\unit{GeV}$, which sets
the inverse coupling of the axion to the SM fields.  Thus the
current data suggests that the axion is basically `invisible' and
very light, with $m_a \lsim 1\unit{eV}$.  Apart from the above 
considerations, axion-type particles are also ubiquitous in string theory.

\begin{figure}
\includegraphics[width=\columnwidth]{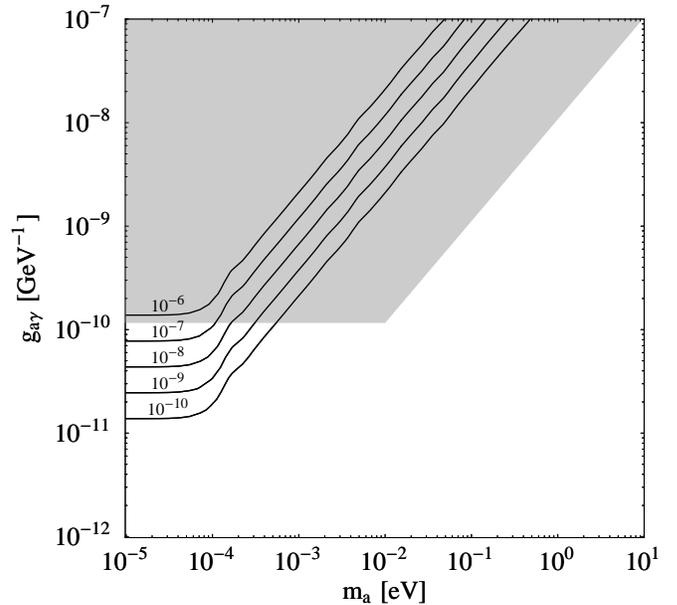}
\caption{\label{fig:sens} X-ray flux from axion conversion in units of
  $\mathrm{photons}\,\mathrm{cm}^{-2}\,\mathrm{s}^{-1}$
in the range from $1-10\,\mathrm{keV}$ for an orbit of
$600\,\mathrm{km}$. The uppermost line corresponds to
the sensitivity of existing satellites, whereas the lowermost line
could be obtained by future missions.
The shaded area schematically depicts the CAST $95\%$ C.L. excluded
region~\cite{Andriamonje:2004hi}.}
\end{figure}

The coupling of the axion to photons is given by
\beq
{\cal
L}_{a\gamma} = -\frac{a}{4 M} F_{\mu \nu} {\tilde F}^{\mu\nu},
\label{La}
\eeq
where $M \sim (\pi/\alpha)f_a$, $\alpha \simeq
1/137$ is the fine structure constant, and $F_{\mu\nu}$ is the
electromagnetic field strength.  The interaction in (\ref{La})
makes it possible for hot plasmas, like the Sun, to emit a flux of
axions through the Primakoff process \cite{Primakoff}.  This same
interaction has also led to experimental proposals
\cite{Sikivie:1983ip} for detecting the axion through its
conversion to photons in external magnetic fields. Various
experimental bounds, most recent of which is set by the CAST
experiment \cite{Andriamonje:2004hi},
suggest that $g_{a\gamma} \equiv
M^{-1} \lsim 10^{-10}\unit{GeV}^{-1}$, as shown in
Fig.~(\ref{fig:sens}). Here we note that some cosmological
considerations related to overclosure of the universe suggest a
lower bound $g_{a\gamma} \gsim
10^{-15}\,\mathrm{GeV}^{-1}$~\cite{Preskill:1982cy}. For a review
of different bounds on axion couplings, see
Ref.~\cite{Eidelman:2004wy}.

In what follows, we propose a new approach
for detecting solar axions, via
their conversion to an x-ray flux
in the magnetosphere near a planet, using a
detector in orbit
\footnote{With small modifications,
our calculations may also be applied to a light
$CP$ conserving scalar $\varphi$, with a coupling to the
electromagnetic field
of the form $\varphi F_{\mu \nu} F^{\mu\nu}$.}.
The possibility of using planetary 
magnetic fields as a conversion region for 
high energy cosmic axions was discussed in Ref.~\cite{Zioutas:1998ra}.  
We take Earth as our
reference example and consider the upward flux of axions going
through the Earth and exiting on the night side
\footnote{Higher sensitivities can be reached for
Jupiter, since its larger magnetic field overcompensates for
the drop in the solar axion flux at the Jovian orbit.}.  This
setup provides an effective way of removing the solar x-ray background.
The radius of the
Earth is $\re \approx 6.4\times10^3\unit{km}$ and its magnetic field is well
approximated by a dipole for distances less than $1000\,\mathrm{km}$
above the surface. The field strength is
$\be \simeq 3\times10^{-5}\unit{T}$ at the equator and it drops
as $1/r^3$~\cite{Landolt}.  However, over distances $L \ll \re$, we may assume
$\be =const$.  We will later show that we are interested in $L
< 1000\unit{km}$, for which this is a valid assumption.

The Earth's atmosphere is mostly composed of nitrogen and oxygen.
Solar axions have an average energy $\vev{\omega_a} \simeq
4\unit{keV}$~\cite{vanBibber:1988ge},
which upon conversion in the magnetosphere will turn
into x-ray photons of the same energy.  The absorption length
$\lambda_{\rm x}$ for 4-keV x-rays in the Earth's atmosphere is
about $10\unit{cm}$ at sea-level~\cite{X-ray}.  However, at an altitude of $150\unit{km}$,
atmospheric pressure falls to about $10^{-10}\unit{atm}$.  For an ideal
gas, density is proportional to pressure.
Since the absorption length of a photon is
inversely proportional to the density of scatterers,
it follows that the x-ray absorption
length scales as the inverse of
pressure.  Thus, above an altitude of $150\unit{km}$, $\lambda_{\rm x}
\sim 10^6\,{\rm km} \sim (10^{-16}{\unit{eV}})^{-1}$ is a lower bound.
Since we will mostly consider $L \gg 150\unit{km}$, this lower bound
holds over distances of interest to us.

The axion to photon conversion probability,
in a transverse magnetic field
of strength $B$, is given by~\cite{Raffelt:1987im,vanBibber:1988ge}
\beq
p_\gamma(L) = \left(\frac{B}{2 M}\right)^2
\left[\frac{1 + e^{-\Gamma L} - 2 e^{-\Gamma L/2}
\cos(q L)}{q^2 + (\Gamma^2/4)}\right],
\label{p1}
\eeq
where $L$ is the path length traveled by
the axion, $\Gamma=\lambda_{\rm x}^{-1}$
is set by the absorption length, and
$\ell=2\pi/q$ is the oscillation length.  Here, we have
\beq
q = \left|m_\gamma^2 - m_a^2\right|/(2 \omega_a),
\label{q}
\eeq
with $m_\gamma$ the plasma mass of the photon.
Since the atmospheric
pressure drops to below $10^{-10}\unit{atm}$
at altitudes small compared to $L\sim 1000\unit{km}$, it is safe to
ignore $m_\gamma$.  To see this, first note that
the largest electron binding energies
for nitrogen and oxygen are well below $1\unit{keV}$~\cite{X-ray},
therefore 
the energy dependence of $m_\gamma^2$ is not
complicated by resonant scattering for x-rays of energy near $4\unit{keV}$.  Thus,
$m_\gamma^2$ is proportional to the density and hence to
the pressure of the gas.  At $1\unit{atm}$,
the plasma mass of 4-keV x-rays
is $m_\gamma\sim 1\unit{eV}$.  This suggests that above an altitude of
$150\unit{km}$, where the pressure falls below
$10^{-10}\unit{atm}$, $m_\gamma$ is less than $\sim 10^{-5}\unit{eV}$,
and hence below the range we consider for $m_a$,
as seen from Fig.~\ref{fig:sens}.
Given an oscillation length
$L \sim 1000\unit{km}$,
we are sensitive to $m_a \sim 10^{-4}\unit{eV}$.  For this value of
$m_a$, we get $q \sim 10^{-12}\unit{eV} \gg \Gamma$,
where $\Gamma = \lambda_{\rm x}^{-1}
\lsim 10^{-16}\unit{eV}$; we may safely ignore
$\Gamma$ in our treatment.

Hence, we can write Eq.(\ref{p1}) as
\beq
p_\gamma(L) = 2\left(\frac{B}{2 M}\right)^2
\left[\frac{1 -
\cos(q L)}{q^2}\right],
\label{p2}
\eeq
For $m_a\leq10^{-4}\unit{eV}$, $M = 10^{10}\unit{GeV}$, $B=\be = 3\times 10^{-5}\unit{T}$,
$\omega = 4\unit{keV}$, and
$L = L_\oplus = \pi/q_{\rm max} \simeq 600\,\mathrm{km}$, we then find
$p_\gamma(L_\oplus) \approx 10^{-18}$.  Here, $q_{\rm max}$ 
corresponds to $m_a = 10^{-4}\unit{eV}$.  Given that the flux of
solar axions at Earth is~\cite{vanBibber:1988ge,Andriamonje:2004hi}
\beq
\Phi_a =
3.67 \times 10^{11} (10^{10}\,{\rm GeV}/M)^2\,{\rm axions}\,{\rm cm}^{-2}
\,{\rm s}^{-1},
\label{Phia}
\eeq
the expected flux of x-rays at an altitude of about $L_\oplus$ near
the equator, for $M = 10^{10}\unit{GeV}$,
is
\beq
\Phi_\gamma(L_\oplus)
\approx 4\times 10^{-7}\,{\rm photons}\,{\rm cm}^{-2}\,{\rm s}^{-1}.
\label{Phigam}
\eeq
Then, for an effective detector area
$A \sim 10^4\unit{cm}^2$ and running time
$\delta t \sim 10^7\unit{s}$,
the number of x-ray photons observed is
$N_\gamma \sim 10^4$.  The signal
decreases as $g_{a\gamma}^{4}$, and thus
this number of events can constrain $g_{a\gamma}$ to near
$10^{-11}\,\mathrm{GeV}^{-1}$.  Hence,
in the regime $m_a \lsim 10^{-4}\unit{eV}$,
the low-Earth orbit observations
can be sensitive to couplings
roughly one order of magnitude smaller than
the current laboratory limits. Figure~\ref{fig:sens} shows the
expected x-ray flux at $L_\oplus=600\,\mathrm{km}$
as a function of $m_a$
and $g_{a\gamma}$. For
this plot we integrated the conversion probability as given in
Eq.~(\ref{p2}) folded with solar axion
spectrum~\cite{vanBibber:1988ge,Andriamonje:2004hi}
over axion energies from $1-10\,\mathrm{keV}$.

Here, we would like to note that matching $m_a$ and $m_\gamma$
results in resonant axion-photon conversion, which in principle
can enhance the signal for $m_a \neq 0$
\cite{vanBibber:1988ge,Raffelt:1987im}. For the solar axions we
consider, there could be a similar effect in the Earth's
atmosphere. However, it turns out that the thickness of any such
resonant layer is always small compared to the oscillation length
of axions and therefore no enhancement results.

It is instructive to have a simple quantitative comparison between
our space-based method and that of laboratory experiments like
CAST.  Here, we will estimate
the figure of merit ${\cal F}$
for each approach.  We note that in the low mass region of interest
to us, the conversion probability scales as $(B L)^2$.  Therefore,
we define ${\cal F} \equiv BL$.
For the CAST experiment, the
transverse magnetic field $B \approx 10\unit{T}$ and the length of
the magnetized region $L \approx 10\unit{m}$ \cite{Andriamonje:2004hi}.
Thus, ${\cal F}({\rm CAST}) \approx 100\unit{T}\unit{m}$.
For the technique presented in this paper, we have $\be \approx 3\times
10^{-5}\unit{T}$ and $L_\oplus \approx 600\unit{km}$, hence
${\cal F}_\oplus \approx 18\unit{T}\unit{m}$.  We see that our approach has a
figure of merit 5 times smaller than
that of the CAST experiment.  However,
the effective magnetized cross-sectional area of the
CAST experiment is about $14\unit{cm}^2$ \cite{Andriamonje:2004hi}.
In our case, since the magnetized
region has a size of order $1000\unit{km}$, the cross sectional
area is only limited by the detector size, which is of order $10^4\unit{cm}^2$.
Hence, given
the same length of time, the space-based technique detailed above has higher
sensitivity than the CAST experiment, for $m_a\lsim 10^{-4}\unit{eV}$.

The above exposition assumes that there are no backgrounds to the
measurement, which in any realistic experiment is not the case. Since
it is difficult to reliably estimate the background from first
principles, it is useful to look at actual x-ray experiments.

One example is the `Rossi x-ray timing explorer' (RXTE) launched in the
mid 90's~\cite{rxte}. It has an effective x-ray collection area of $\sim
7\,000\,\mathrm{cm}^2$ and is sensitive over $1-50\,\mathrm{keV}$. Its
angular resolution is roughly $0.5^\circ$. It orbits the Earth at a
height of $\sim 600\,\mathrm{km}$. During slews and calibration from
1996 till 1999 it has
acquired at least $25\,000\,\mathrm{s}$ of viewing the night side of
the Earth. The nominal background when watching the blank
sky\footnote{This includes the contribution from the diffuse cosmic
  x-ray background, therefore the sensitivity given here is an
upper bound.} is
about 3 counts per second in the energy range from
$2-10\,\mathrm{keV}$ over the whole effective
area~\cite{Revnivtsev:2003wm}.
Thus there are
$7.5\times10^4$ background events in $25\,000\,\mathrm{s}$, whose
fluctuations are given by $\sqrt{7.5\times10^4}\sim270$; dividing this
by the exposure and area gives a sensitivity
of $1.5\times10^{-6}\,\mathrm{photons}\,\mathrm{cm}^{-2}\,\mathrm{s}^{-1}$.
Although this experiment was not designed to perform our type of
observation, this level of sensitivity would allow to probe axion couplings
$g_{a\gamma}$ of the order $10^{-10}\,\mathrm{GeV}^{-1}$, which is shown
in Fig.~\ref{fig:sens}.

Another example is the LOBSTER~\cite{lobster} experiment planned
to fly on the International Space Station, which has an orbit of
$350\,\mathrm{km}$. It will have a total x-ray collection area of
$\sim 10^4\,\mathrm{cm}^2$, its angular resolution is
approximately $3'$ and it is sensitive from
$0.5-3.5\,\mathrm{keV}$. It has a background rate per pixel (the
solar core where axions are produced covers approximately 1 pixel)
of $10^{-5}\,\mathrm{s}^{-1}$. Thus, in $10^7\,\mathrm{s}$ of
observation they have $100$ background events and the background
fluctuation is $\sqrt{100}=10$ events.  Therefore, we have a
sensitivity of
$10/10^4/10^7=10^{-10}\,\mathrm{photons}\,\mathrm{cm}^{-2}\,\mathrm{s}^{-1}$,
which in turn, according to Fig.~\ref{fig:sens}, gives a limit on
$g_{a\gamma}$ of order $10^{-11}\,\mathrm{GeV}^{-1}$.

Given the above considerations, existing or planned orbital x-ray
telescopes can be used for solar axion detection, using the
method proposed here.  Therefore, a specialized and
dedicated mission is not required to implement our approach.  In fact,
some of the existing x-ray data may already contain the information
required for competitive or even better sensitivities than those
obtained from current laboratory experiments.
This depends on how much of the present
x-ray calibration data
has been obtained while the telescope was
pointed at the core of the Sun, on the dark side of the Earth.

To summarize, in this Letter,
we have proposed a new technique for detecting solar axions, using
Earth's magnetosphere.  We have shown that given the large magnetized volume
around the Earth, conversion of axions, with mass $m_a \lsim 10^{-4}\unit{eV}$, into
x-rays of average energy $\vev{\omega} \simeq 4\unit{keV}$ will be measurable by an
x-ray telescope, in a low-Earth orbit.
A key ingredient of our proposal is
to use the Earth as an x-ray shield and look for axions coming through the Earth
on the night side.  This effectively removes the solar x-ray background.
Thus, observation of x-rays, with a thermal energy
distribution peaked at approximately $4\unit{keV}$, on the night side of the Earth is a
distinct signature of solar axions in our proposal. Moreover these
x-rays would only come from the center of the Sun, which subtends
approximately $3'$ and there would be an orbital variation with
magnetic field strength and an annual modulation by the Sun-Earth
distance.  Considering the well-established framework of the
solar model, it would be extremely difficult to come up with an
alternative explanation of
all these signatures.
Therefore, our method
can achieve an unambiguous detection of solar axions.
We estimate that, in a one-year run,
this technique will gain sensitivity to axion-photon couplings of order
$g_{a \gamma} \sim 10^{-11}\unit{GeV}^{-1}$,
about an order of magnitude beyond current laboratory limits.
We conclude that, for solar axions in the regime considered here, our approach
will probe axion-photon couplings beyond the reach of foreseeable laboratory
experiments.

\acknowledgments

It is a pleasure to thank D. Chung and G. Raffelt for a critical
reading of the manuscript and V. Barger for discussions
and helpful comments.
This work was supported in part by the United States Department of
Energy under Grant Contract No. DE-FG02-95ER40896.  H.D. was also
supported in part by the P.A.M. Dirac Fellowship, awarded by the
Department of Physics at the University of Wisconsin-Madison.


\end{document}